# A Pilot Case Study on Innovative Behaviour: Lessons Learned and Directions for Future Work

Cleviton V. F. Monteiro
Universidade Federal Rural de Pernambuco
Rua Dom Manoel de Medeiros, s/n
Dois Irmãos, Recife-PE, Brazil
+55 81 33206590
cleviton@gmail.com

Fabio Q. B. da Silva
Universidade Federal de Pernambuco
Av. Jornalista Anibal Fernandes, s/n
Cidade Universitária, Recife-PE, Brazil
+55 81 21268430
fabio@cin.ufpe.br

Luiz Fernando Capretz
University of Western Ontario
Thompson Engineering Building
N6A 5B9, London, Canada
+1 519-661-2111
lcapretz@uwo.ca

## ABSTRACT

**Context**: A case study is a powerful research strategy for investigating complex social-technical and managerial phenomena in real life settings. However, when the phenomenon has not been fully discovered or understood, pilot case studies are important to refine the research problem, the research variables, and the case study design before launching a full-scale investigation. The role of pilot case studies has not been fully addressed in empirical software engineering research literature. **Objective**: To explore the use of pilot case studies in the design of full-scale case studies, and to report the main lessons learned from an industrial pilot study. **Method**: We designed and conducted an exploratory case study to identify new relevant research variables that influence the innovative behaviour of software engineers in the industrial setting and to refine the full-scale case study design for the next phase of our research. **Results**: The use of a pilot case study identified several important research variables that were missing in the initial framework. The pilot study also supported a more sophisticated case study design, which was used to guide a full-scale study. **Conclusions**: When a research topic has not been fully discovered or understood, it is difficult to create a case study design that covers the relevant research variables and their potential relationships. Conducting a full-scale case study using an untested case design can lead to waste of resources and time if the design has to be reworked during the study. In these situations, the use of pilot case studies can significantly improve the case study design.

## 1. INTRODUCTION

Innovative behaviour is a multidimensional construct defined as "*the intentional generation, promotion, and realization of new ideas within a work role, work group, or organization in order to benefit role performance, a group, or an organization*" [6]. Examples of such behaviour include the suggestion of new products and processes, the adoption of new technologies, and the application of new working methods. In our research about human factors in industrial software engineering practice, we observed and catalogued several examples of innovative behaviour exhibited by software engineers with positive impacts at the individual, team, and organizational levels. The benefits of innovative behaviour in practice motivated us to investigate which factors foster or suppress this behaviour at the individual, group, and organizational levels.

As a starting point, we conducted an *ad hoc* literature review covering innovative behaviour models from several fields. The findings showed almost no study focusing on software engineers and software organizations. Further, the studies from other areas showed no consensus on a theory, and their results were impossible to be compared. Several authors have argued that case study is a suitable choice of research method to early exploratory investigations of a phenomenon and to build "provisional" theories when none is available or widely accepted [5][10][17]. Christie et al. [5] suggested the use of pilot case studies in such contexts to refine the research problem and variables, and the case study design as a whole, before committing resources to full-scale studies. However, as far as we are aware, there is no published example of this use of pilot case studies in software engineering.

Therefore, in this article we describe how a full-scale case study design was built from the results of a pilot case study. We then discuss some lessons learned emphasizing the role of pilot studies in the construction of more robust case study designs.

## 2. BACKGROUND

The background related to innovative behaviour and the three existing models is summarized in this section.

### 2.1 Innovative Behaviour

Innovative behaviour is viewed as a multistage process [18] that starts with an individual creating and proposing a new (potentially useful) idea. Then, this individual promotes the idea to gain support from colleagues, managers, or sponsors. Finally, the idea can be operationalized with the production of a prototype, a proof, a concept, or the use of a new technology within a software project.

### 2.2 Innovative Behaviour Models

The innovative behaviour phenomenon has been studied in several areas [1][18][12], but we could not find any study reporting results from the software development industry. Using findings from diverse fields, we found three models that attempt to explain the antecedents of innovative behaviour [1][18][21]. In the model proposed by Åmo [1], the individual innovative behaviour is positively influenced by 12 factors, which can be grouped into four categories:

- *Characteristics of the organization*: expressed strategy and size of the organization;
- *Characteristics of the intersection between employee and employer*: hierarchy, organization desire expressed by management, culture of the work group, and level of specialization in job function;

- *Characteristics of the actual individual*: proactive personality, intrapreneurial personality, eagerness for learning, and age;
- *Characteristics of the innovation itself*: embedded learning potential and fitness with organizational goals.

Scott and Bruce [18] proposed that innovative behaviour is influenced by: leader role expectations, leader-member exchange, the individual intuitive problem-solving style, the individual systematic problem-solving style, the individual career stage, and the climate for innovation. However, the potentially complex interactions among those factors are not described in their model.

West [21] proposed that group creativity and behaviour towards implementation are influenced by a composition of four interacting factors: group task characteristics, group knowledge diversity and skills, integrating group processes, and external demands. In particular, external demand is a new element to be considered in the study of work group creativity and innovation, and it has not been previously addressed in the literature.

Analysing these three models, we observed the following gaps:

- They propose different variables to explain the innovative behaviour, with few overlaps, which makes the models almost impossible to compare.
- Two of them [1][18] studied the innovative behaviour phenomenon at the individual level, while West [21] studied it at the group level. This also makes it difficult to compare the models.

These studies were performed in several different industries but none of them focused on software organizations. According to Hackman [11], the relationships among factors that explain individual behaviour during teamwork seem to depend substantially on the properties of the group task being performed. This reinforces the need to study innovative behaviour in the software industry.

## 3. THE PILOT CASE STUDY

Several authors suggested the use of a pilot case study when the phenomenon of interest is not fully discovered and understood [5][10][17]. To design our pilot case study, we followed the method proposed by Eisenhardt [10]. The full report of these results and the detailed case study protocol is presented elsewhere [16].

### 3.1 Getting Started

We started with the definition of our research question: *How is innovative behaviour of software engineers supported or supressed in software development industrial practice?*

Then, we built the pilot case study design. We chose the *software engineer professional* as the unit of analysis, because the research question is directly related to the expression of the phenomenon at the individual level. In addition, the design also had to deal with the contextual factors related to the unit of analysis. In this case, contextual factors were considered at three levels (based on the models discussed in Section 2.2): the software team, the team leader or project manager, and the organization itself.

The existing literature about innovative behaviour did not provide fully accepted and consistent theories or models that supported the identification of exactly which variables to observe, control, and vary with respect to three levels of contextual factors. Therefore, we needed a flexible design to allow the identification of new relevant variables.

Figure 1 depicts the pilot case study design. We investigated a single software organization and studied individuals from two different projects, with different team leaders. With this design we obtained variability of teams and leaders, while keeping the organizational context fixed. To obtain variability at the individual level, we used the criterion explained in Section 3.2.

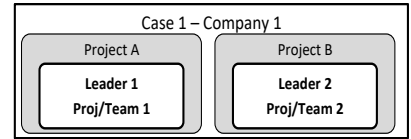

**Figure 1. Pilot Case Study**

### 3.2 Selection of participants

We investigated individuals with low, medium, and high innovative behaviour to compare their behaviours and what influenced them. To select the participants, the project manager of each project classified the team members according to the frequency they behaved innovatively, following the innovative behaviour definition that we presented to them. The project managers were also interviewed to allow data triangulation. The limitations regarding this method of participant selection are discussed below.

### 3.3 Data Collection

We used interviews and observation to collect data from participants. Semi-structured interviews were performed with software development team members and project managers. The two interview guides (for team members and managers) were composed of open questions combined with probing questions. Both guides were piloted with individuals from a company that did not participate in the study. The audio of all the interviews was recorded and transcribed verbatim.

Observation was chosen to allow the researchers to monitor behaviour and interaction among team members that could not be obtained from interviews [15]. The observations happened during the project meetings and focused on identifying idea proposal, and the past or present implementation of an idea proposed by the team members.

### 3.4 Data analysis

Data analysis was performed in parallel with data collection, in incremental and iterative steps. We used coding techniques from grounded Theory [15] to code, categorize, and synthesize data. We used QSR NVivo[1] to support the data analysis and synthesis.

Data analysis began with open coding of the transcripts. Post-formed codes were constructed as the coding progressed and were attached to particular pieces of the text with the support of NVivo. An example of a complete code is *C1PATM2_No financial rewards*, which means that the evidence points to the code "No financial rewards" and was collected from the interview of team member 2, who worked on project A in Company 1.

Then, we grouped the codes into categories that affect innovative behaviour. As the process of data analysis progressed, we built the interacting effects of these factors, expressed as propositions, and created a model that described the innovative behaviour of individuals in this organization.

---

[1] www.qsrinternational.com/products_nvivo.aspx

## 3.5 Enfolding the Literature

We then looked at the literature to sharpen construct definitions and generalizability, and raise the theoretical level. In addition to the literature review discussed in Section 2, a supplementary literature review was performed after the pilot case study. While the pilot case study provided new variables to be investigated, this review focus was to provide a theoretical foundation to the findings and to support the refinement of the case study design (see in Section 4).

## 3.6 Model Development

We synthesized the findings from the coding process to create a model that represent the relationships among factors related to innovative behaviour. The model, called Initial Innovative Behaviour Model for Software (IBMSW-*i*), is shown in Figure 2 and described in detail by Monteiro et al. [16].

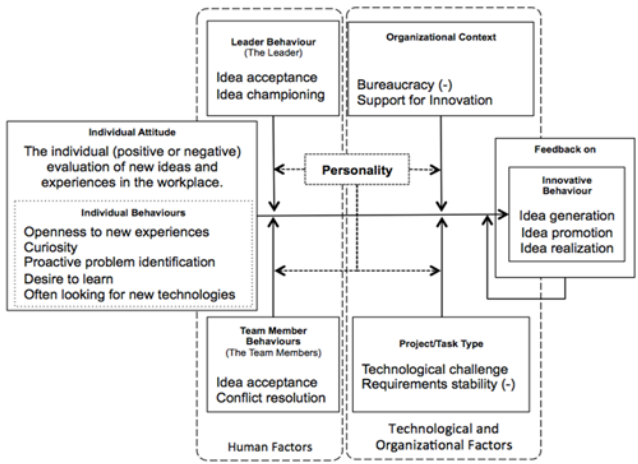

**Figure 2. The IBMSW-*i***

The core element of this model, which is novel in the literature about innovative behaviour, is the direct influence of individual attitude on the expression of innovative behaviour. External signs of this attitude are behaviours related to curiosity, desire to learn, proactivity, etc. The expression of innovative behaviour is also indirectly influenced by situational or contextual factors in the workplace. These factors create conditions that will be perceived and interpreted by the individuals, and will, in turn, moderate the expression of innovative behaviour at the individual level. We grouped these factors into two higher-level categories: those containing **Human Factors** and those containing **Technological and Organizational factors**. From the findings related to the Human Factors category, we built a hypothesis:

> **Hypothesis** – *The relationship with peers (team members and leaders) at the workplace will indirectly affect the expression of innovative behaviour through the creation of (favourable or unfavourable) working conditions for idea proposition, promotion, and implementation.*

The organization as a whole also influences the expression of innovative behaviour. The organizational factors and the uncertainty levels of the tasks related to technological aspects are likely to be interrelated, as expressed in this hypothesis:

> **Hypothesis** – *Higher levels of task uncertainty (requirements flexibility and technological challenge) in the presence of support for innovation and low bureaucracy in the organization will indirectly affect innovative behaviour through its moderating effect on the relationship between individual attitude and individual innovative behaviour.*

We also postulate that individuals would react differently to the situational factors depending on their personality traits. Further, the expression of innovative behaviour evolves over time, contingent on the feedback received.

## 4. THE REFINED CASE STUDY DESIGN

Christie et al. [5] suggested the use of pilot case studies to refine the research problem, the research variables, and the case study design as a whole, before committing resources to full-scale studies. Similarly, Runeson emphasized that "a pilot case [can be used] to explore the phenomenon under study, and the following cases may be used for more in-depth investigations" [17].

We used the IBMSW-i model to guide the refinement of a full case study design. We started by selecting factors from the model to guide the sampling of projects, teams, and individuals. We then investigated theories to raise the theoretical level of the factors selected and to provide data collection instruments (Section 4.1). Finally, we created the new case study design (Section 4.2).

## 4.1 Selecting Factors

We wanted to select factors from the IBMSW-*i* to improve the design of the case study, in particular regarding sampling the projects and participants. In the pilot case, our design selected projects and participants from a single company. We decided to keep this design choice because we still do not know enough about which organizational characteristics would be important to guide the selection of new organizations.

As we looked at the individual behaviour expressed in the context of a software team, we decided to use the Project Type and the **Leader's Behaviour** to guide the sampling of projects. To do this, we needed two operational definitions that could be used to distinguish styles of leadership and types of projects.

Regarding the sampling of participants in each project, we decided to look for a suitable **operational definition of innovative behaviour**. Finally, to be able to investigate the moderating effects of personality, we also decided to select a **personality test** to be used.

### 4.1.1 Project Type

The pilot case study was conducted with participants of software development projects. The results from the pilot study indicated that the *requirements stability* in the project and its *technological challenges* influenced individual innovative behaviour. We needed a classification model that could distinguish the projects regarding their requirements stability and technological challenges. The Three Horizons Model [20] (Figure 3) proposes a classification scheme according to the levels of two orthogonal uncertainties faced by projects: technological uncertainty and market uncertainty. Technological uncertainty is defined by the organization's ability to overcome the technical difficulties of an opportunity. In turn, market uncertainty is defined by the organization's ability to understand and address the needs of a group of customers.

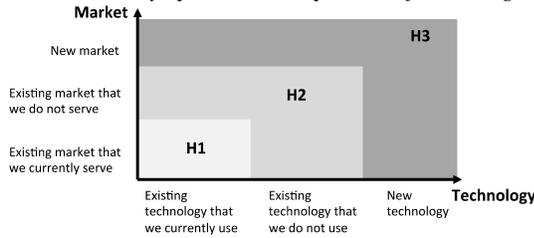

**Figure 3. Three Horizons Model**

Using the technological and market uncertainties as an axis, the model defined three spaces of innovation:

- Horizon 1 (H1): projects that involve mature technologies and that are targeted to the markets already served by the organization are classified as H1. In this horizon, the risk is small and the innovations are marginally incremental.
- Horizon 2 (H2): projects that involve technologies that are new to the organization and/or that are targeted to a market that the company has not yet explored are level H2. Such technologies already exist and are available, but they are not dominated by the organization. In H2, there are relative uncertainties and projects with a moderate level of innovation.
- Horizon 3 (H3): projects that involve emerging technologies and/or are targeted to a market that does not yet exist (are untapped by any other organization) are level H3. Such technologies are still in development or have been used in an experimental way. H3 projects have a high level of uncertainty and can provide the highest opportunities for innovation.

In this model, more uncertainty (technical or market) is likely to be related to less stable requirements and/or more technical challenges and, consequently, more space to change.

In the pilot case study, both projects were H1 and the influence of the requirements stability and technological challenge emerged when comparing the current projects with projects in which the participants had worked in the past. Thus, using the Three Horizons Model it will be possible to select projects from different horizons in future case studies to obtain greater variability of requirements stability and technological challenge.

### 4.1.2 Leadership Style

The variables related to the leader's behaviour found in the pilot case study were the **leader acceptance to ideas**, the **feedback**, and the **autonomy** provided to the individuals to perform their tasks. However, we could not find studies investigating the influence of these specific variables on the individual innovative behaviour. Thus, we used the results of a broader search that was performed using a systematic literature review (SLR) [9] to compile the leadership influence on individual innovative behaviour. The research questions of the SLR guided our analysis:

*RQ1. How do leaders influence the innovative behaviour of individuals?*

*RQ1.1. Which of the leader factors are most studied?*

In this SLR, we found two theories connected to the variables related to leader behaviour found in the pilot case study: the transformational and transactional leadership styles.

The transformational leader "*raises associates level of awareness of the importance of achieving valued outcomes and the strategies for reaching them*" [4]. They also encourage followers to transcend their self-interest for the sake of the team or organization. Further, they encourage followers to raise their needs in areas such as achievement, autonomy, and affiliation [4]. Burns was the precursor of the transformational leadership theory and Bass and Avolio [2] helped it to evolve.

In turn, the transactional leadership style [3] builds the foundation for relationships between leaders and followers in terms of clarifying responsibilities, specifying expectations and task requirements, negotiating contracts, and providing recognition and rewards in exchange for the expected performance [14]. The transactional leader usually operates to guarantee that subordinates will work according the existing culture. Such leaders pay close attention to deviations, irregularities, and mistakes in order to take action and make corrections.

Thus, considering these leadership styles, we related them to the pilot case study variables using the following rationale. Transformational leaders stimulate the individual using influence and motivate them to engage in actions to promote change. On the one hand, the transactional leader uses the explicit task definition to control and measure performance. Relating this to the leadership variables we found, we saw that when the leader provided low or no autonomy to the individuals and did not accept changes, there was strong control over their actions and tasks. This situation is directly related to the characteristics of the transactional leadership style. On the other hand, the transformational leader is related to higher levels of delegation, autonomy, and openness to change.

Using this rationale, we can use the transformational and transactional leadership style theories to raise the theoretical level (use more precise definition of the constructs) of the category **Leader's Attitude**.

### 4.1.3 Operational Definition of Innovative Behaviour

To overcome the limitation related to measuring participant innovative behaviour, we decided to use the operational definition suggested by Scott and Bruce [18]. They proposed a six-item scale that should be rated by the manager for each team member. Examples of such items are: "generates creative ideas" and "investigates and secures funds needed to implement new ideas." The responses should be examined using a five-point Likert scale, ranging from *not at all* to *an exceptional degree*.

### 4.1.4 Personality

The pilot case study showed that some individual's characteristics explained her behaviour towards proposing ideas and implementing them. The following excerpts were extracted from different professionals and exemplify this finding.

*"Once I learn how it works it becomes boring. So I have this intrinsic need to try to make things... to eliminate as much boring as I can, so I can focus on the interesting parts." [C1PATM1]*

*"Unless I see a problem, or try to resolve a situation, I would not have an incentive to research on new idea or new way to do things. But they are totally personal things." [C1PATM3]*

Therefore, the psychology literature was analysed with the aim to understand the influence of the individual personality. Among the various theoretical foundations, traits, and types theories we looked for those that are most used in organizational psychology and in the studies about personality in software engineering [8]. In particular, the Five-Factor Model (FFM) [7] has been used both in software engineering [8] and creativity [13] researches. Thus, we decided to use the FFM [7] to guide our understanding of the influences of personality on innovative behaviour.

## 4.2 Refining the Case Study Design

Our goal, after the pilot study, was to refine the initial design to achieve variations on project type (Three Horizons Model) and the leadership styles of project managers. Figure 4 illustrates the refined case design after adding these two criteria to sample projects. This generic design must be instantiated by choosing the number of projects in each quadrant. In general, we do not know this number up front because the size of the sample in qualitative studies is often defined as the study progresses [14].

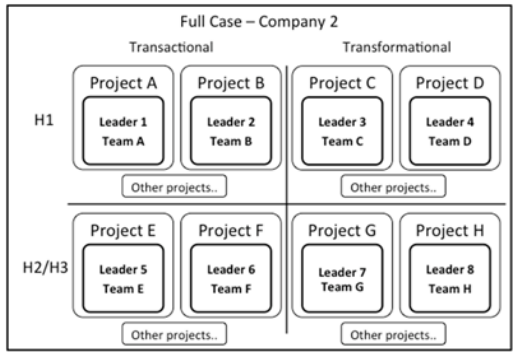

**Figure 4. Design of the full case study**

In the new design, we propose the sampling of projects that are managed by transactional managers and by transformational managers. This shall allow a comparison of the influence performed by different leadership styles on individual innovative behaviour. To assess the leadership style of project managers, the MLQ questionnaire developed by Bass and Avolio [2] is a suitable operationalization of the construct.

The refined design also separates projects according to uncertainty horizons. To operationalize this factor, we split the projects into two groups: one with horizon H1 projects and the other with horizons H2/H3. This design allowed a comparison of individual innovative behaviour when working on projects with low uncertainty (stable requirements and low technological challenges) and medium to high uncertainty (unstable and open requirements and medium/high technological challenges).

In addition, two instruments also enhanced the data collection process. First, we included the administration of the FFM questionnaire [7] to provide information about participant scores on each personality trait. Second, we included Scott and Bruce's [18] instrument to evaluate the innovative behaviour level of participants. The goal was to use this additional information in two different ways. Quantitatively, it could be used to identify the existence of correlations between the individual's personality traits and her innovative behaviour. Qualitatively, it could be used to explain specific individual behaviour according to the scores on each personality trait.

It is important to highlight that this refined design was created to increase the diversity regarding the factors uncovered in the pilot study. We did not have as an objective to control or to manipulate variables as performed on controlled experiments.

## 4.3 Lessons Learned

Before the pilot case study, we did not have established models of theories to guide our investigation. The existing studies in other areas were not conclusive, were mostly difficult to compare, and addressed tasks and jobs substantially different from those found in the software industry. Therefore, we could only design a simple case study design to explore the phenomenon and uncover new factors and potential relationships. We believe that the following lessons are important for researchers facing similar situations:

- *Understand the dual role of the pilot case study*: a pilot case study can indeed support the development of initial, provisional theories, such as the IBMSW-*i*, when no one exists. It can also be instrumental in uncovering new factors or design issues not previously addressed. Therefore, pilot case studies can produce results at the substantive and the methodological levels of the Research Path Scheme [19].
- *Do not use pre-defined models or theories*: consistent with an interpretive or constructivist stance, try to avoid the potential biases of entering the field with pre-defined models or theories guiding your investigation. This could blind you to new factors not addressed in these models or theories.
- *Keep the design simple*: because we have limited knowledge of the phenomenon in the context of study, it is important to keep the design of the case study as simple as possible.
- *Collect as much information as possible*: although you need a simple design, it is important to get as much (potentially unrelated) information as possible. Long interview scripts and several hours of observation are important, even though it may lead to large amounts of data and increase the complexity of the data analysis.
- *Do not use pre-formed codes in data analysis*: as in the second point above, try not to use pre-formed codes in your data analysis. Although this type of technique helps in making sense of large amount of data, it can also hide new factors or relationships.
- *Be grounded on the data, but with freedom to create*: bear in mind that your design, by construction, may not support the production of necessary data to uncover all aspects of the phenomenon. Therefore, be faithful to your data but allow yourself to produce inductions and abductions that fill gaps and explain inconsistencies, perhaps in the form of propositions and hypothesis.
- *Use the models or theories to refine the design*: the starting point to refine the case study design is the results of the pilot. Use these results to identify new variables, their operationalization, and more robust sampling strategies.
- *Decrease the breadth of the data collection*: after learning with the pilot case study, you can be less exploratory in your data collection, leaving out information that was not relevant. However, exercise this advice with caution according to your understanding of the phenomenon to not leave out relevant information.
- *Increase the depth of the data collection*: now that you know what matters in your study, collect more in-depth information about the relevant factors. This may include having more questions in qualitative interview scripts, more observation items, and more types of observation, the investigation of documentations, or even adding quantitative data to increase the richness of the interpretations. At this stage, operational definitions of relevant factors should be provided: either developing new instruments when none exists or using available instruments from the literature.

Although we followed the items above in our research, we have not tested them in other cases. Therefore, not all of them may work in specific contexts. We hope that other researchers, performing pilot case studies would share their lessons learned confirming or revising our suggestions.

## 5. CONCLUSION

We presented the results of a pilot case study conducted to identify factors that influence the innovative behaviour of software engineers in practice. From these results, a preliminary model explaining the relationships among these factors was built (IBMSW-i), answering our research question. This result is fully presented by Monteiro et al. [16]

Our model consistently improves and extends existing models from other fields of study with factors specific to the software development practice, such as the role of requirements stability. However, the pilot case study design did not allow full identification of interacting effects between the factors. Further, project type and individual personality were not addressed in the pilot case study. To progress in the study of innovative behaviour, we produced a refined study design to incorporate new factors and their operationalization. This new design should be better equipped to uncover influences of these factors and to increase construct and internal validity of future studies. We intend to use this refined design in other case studies about innovative behaviour in software organizations.

In addition to the results presented, we learned some lessons that we thought would be worthwhile to share with the research community. Before the execution of the pilot case study we experienced considerable difficulty in designing the full case study. The use of an exploratory pilot case study avoided the development of full-scale case study based on variables that were not important to explain the phenomena at hand, removing the waste of resources and time with rework due to the usage of wrong premises in the design. In addition, the supplementary literature review was important to improve the final design. It also provided tested instruments to improve data collection.

We believe that our results can be used at the substantive and methodological level, as proposed in the Research Path Schema [19]. The IBMSW-*i*, at the substantive level, provides a novel understanding of innovative behaviour in software engineering that can be tested in other contexts. The new refined design (and the lessons learned in its construction) contributes to the methodological level and to researchers performing studies with similar characteristics.

## 6. ACKNOWLEDGMENTS

Fabio Q. B. da Silva holds a research grant from the Brazilian National Research Council (CNPq), process #314523/2009-0. Cleviton Monteiro received a scholarship from CAPES and the ELAP award.

## 7. REFERENCES

[1] Åmo, B. 2005. Employee innovation behavior. Bodø Graduate *School of Business*, Bodø:Norway

[2] Bass, B. M.; Avolio, B. J. MLQ, Multifactor Leadership Questionnaire. *Mind Garden*, Redwood City, CA, 1995.

[3] Bass, B. M. Leadership and performance beyond expectations. *Collier Macmillan*, New York, 1985.

[4] Burns, J. 1987. Leadership. Harper & Row, New York, NY

[5] Christie, Michael; Rowe, Pat, Perry, Chad, and Chamard, John. 2000. Implementation of Realism in Case Study Research Methodology. In International Council for Small Business, Annual Conference (Brisbane)

[6] Cingöz, A.; Akdogan, A. 2011. An empirical examination of performance and image outcome expectation as determinants of innovative behavior in the workplace. Procedia - Social and Behavioral Sciences, 24, 847–853.

[7] Costa, P. T., Jr.; McCrae, R. R. 1992. The NEO Personality Inventory (NEO PI-R) and NEO Five-Factor Inventory (NEO-FFI) professional manual. *Psychological Assessment Resources*, Odessa, FL

[8] Cruz, S.S.J.O.; da Silva, F.Q.B.; Capretz, L.F. 2015. Forty years of research on personality in software engineering: A mapping study. Computers in Human Behavior, v. 46, p. 94-113, doi:10.1016/j.chb.2014.12.008.

[9] da Silva, F.Q.B.; Monteiro, C.V.F.; dos Santos, I.B.; Capretz, L.F. 2016. How leaders influence followers' innovative behaviour in software development practice. Recife. IEEE Software (*in press*).

[10] Eisenhardt, K. 1989. Building Theories From Case Study Research. *The Academy of Management Review*, 14, 4, 532-550.

[11] Hackman JR. 1987. The design of work teams. In Lorsch J *Handbook of organizational behaviour*, Englewood Cliffs, NJ: Prentice-Hall

[12] Jong, J. P. J.; Den Hartog, D. N. 2007. How leaders influence employees' innovative behavior. *European Journal of Innovation Management*, 10, 41-64.

[13] King, L.; Walker, L.; Broyles, S. 1996. Creativity and the five factor model. *Journal of Research in Personality*, 30, 189–203.

[14] Liu, J.; Liu, X.; Zeng, X. 2011. Does transactional leadership count for team innovativeness? The moderating role of emotional labor and the mediating role of team efficacy. *Journal of Organizational Change Management*, 4, 3, 282-298.

[15] Merriam, B. S. 2009. *Qualitative Research: A Guide to Design and Implementation*. Jossey-Bass, San Francisco.

[16] Monteiro, C. V. F.; da Silva, F. Q. B.; Capretz, L. F. 2016. Innovative Behaviour of Software Engineers: Findings from a Pilot Case Study. In: Proceedings of the 10th ACM/IEEE International Symposium on Empirical Software Engineering and Measurement.

[17] Runeson, P.; Host, M.; Rainer, A.; Regnel, B. 2012. *Case Study Research in Software Engineering: Guidelines and Examples*. Wiley, Hoboken. ISBN 9781118104354.

[18] Scott, S.G.; Bruce, R.A. 1994. Determinants of innovative behavior: a path model of individual innovation in the workplace. *Academy of Management Journal*, 38, 1442-65.

[19] Stol, K.-J.; Fitzgerald, B. 2015. Theory-oriented software engineering. Science of Computer Programming, *Towards general theories of software engineering*, v. 101, p. 79–98

[20] Terwiesch, C.; Ulrich, K. 2009 *Innovation Tournaments: Creating and Selecting Exceptional Opportunities*. Harvard Business School Press.

[21] West, M.A. 2002 Sparkling fountains or stagnant ponds: an integrative model of creativity and innovation implementation in work groups. *Applied Psychology: An International Review*, 51, 3, 355-387.